\documentclass[nojss]{jss}

\usepackage{orcidlink,thumbpdf,lmodern}

\usepackage{thumbpdf}

\usepackage[longend]{algorithm2e}
\RestyleAlgo{ruled}

\usepackage{comment}

\usepackage{amsmath}
\usepackage{amssymb}
\usepackage{pgfplots}
\usepackage{caption}
\usepackage{subcaption}

\usepackage{tikz}
\usetikzlibrary{shapes.geometric}
\usetikzlibrary{arrows.meta}
\usetikzlibrary{positioning}
\usetikzlibrary{calc}

\tikzset{%
    arrow/.style={%
        -{Stealth[length=2mm, width=1.5mm]}
    }
}

\pgfplotsset{compat=1.18}

\usepackage{float}

\usepackage{colonequals}

\newcommand{\Pcf}{\mathsf{Pcf}}
\DeclareMathOperator{\range}{range}

\author{Bj\"{o}rn H. Wehlin~\orcidlink{0000-0002-2114-1000}\\KTH Royal Institute of Technology}
\Plainauthor{Bj\"{o}rn H. Wehlin}

\title{Massively Parallel Computation of Similarity Matrices from Piecewise Constant Invariants}
\Plaintitle{Massively Parallel Computation of Similarity Matrices from Piecewise Constant Invariants}
\Shorttitle{Parallel Computation of Similarity Matrices from Piecewise Invariants}

\Abstract{
  We present a computational framework for piecewise constant functions (PCFs) and use this for several types of computations that are useful in statistics, e.g., averages, similarity matrices, and so on. We give a linear-time, allocation-free algorithm for working with pairs of PCFs at machine precision. From this, we derive algorithms for computing reductions of several PCFs. The algorithms have been implemented in a highly scalable fashion for parallel execution on CPU and, in some cases, (multi-)GPU, and are provided in a \proglang{Python} package. In addition, we provide support for multidimensional arrays of PCFs and vectorized operations on these. As a stress test, we have computed a distance matrix from 500,000 PCFs using 8 GPUs.
}

\Keywords{\pkg{masspcf}, piecewise constant functions, kernels, similarity matrices, \proglang{Python}, \proglang{C++}}
\Plainkeywords{masspcf, piecewise constant functions, kernels, similarity matrices, Python, C++}

\Address{
  Bj\"{o}rn H. Wehlin\\
  Department of Mathematics\\
  KTH Royal Institute of Technology\\
  100 44 Stockholm, Sweden\\
  E-mail: \email{bwehlin@kth.se}
}

\begin{document}



\section[Introduction]{Introduction} \label{sec:intro}



We provide a \proglang{Python} package (\pkg{masspcf}) and accompanying \proglang{C++} library for performing massively parallel computations with piecewise constant functions (PCFs) at machine precision. In rough terms, a PCF is a 1-D function that changes at only a finite number of points. Such functions are abundant in statistics and data science, as well as more broadly within science and engineering.




Our main motivation stems from the rapidly evolving field of Topological Data Analysis (TDA) \citep{frosini1990distance, robins1999towards, edelsbrunner2002topological, ghrist2008barcodes, carlsson2009topology, carlsson2021topological}, where the geometry of finite point clouds is reflected in various discrete invariants. These include, for example, stable rank \citep{scolamiero2017multidimensional, gafvert2017stable, chacholski2020metrics}, Betti curves/sequences \citep{ giusti2015clique, reimann2017cliques, umeda2017time} and Euler characteristic curves \citep{reimann2017cliques, dlotko2022euler}.

We think of the PCF obtained from a point cloud as a time series of numbers. A distinguishing characteristic of such time series is a lack of regularity in the sampled time points. Outside of TDA, a prototypical example of time series of this kind is central bank interest rates. Here, we may want to compare over time the interest rate in one country to that of another country. Obviously, central banks do not move all in unison, and the interest rate is usually fixed for a duration of time. 

Other examples from a more classical statistical perspecitve include change point detection and univariate stochastic processes. One can also think of a histogram as a piecewise constant function.

To do statistics with piecewise constant functions, one should be able to take averages and standard deviations, compute pairwise distances and kernels, and so on. For example, in the field of computational neuroscience, the Earth Mover's Distance has been used to compare spike trains \citep{sihn2019spike}. This involves constructing PCFs and computing their pairwise $L_1$ distances, using the methodology of \citet{cohen1999finding}.

If $f,g \colon [0,\infty) \to \mathbb{R}$ are two PCFs, their $L_p$ ($p \geq 1$) distance, $d_p(f,g)$, and $L_2$ inner product $\langle f, g \rangle$ are defined by
\begin{align*}
    d_p(f,g) \colonequals \left(\int_0^\infty |f(t)-g(t)|^p\right)^{1/p}, ~~~\text{and} ~~~ \langle f, g \rangle \colonequals \int_0^\infty f(t)g(t) \, dt,
\end{align*}
respectively.

We are primarily interested in the case when we have a large number of PCFs and we want to compute a \emph{matrix} of distances (i.e., a distance matrix) or inner products (i.e., a Gram/inner product matrix), and so on.

For a limited number of PCFs, these matrices can be computed without too much effort. However, once there are thousands of PCFs, the number of integrals to compute gets into the millions; and even though these integrals are simple to compute, for 100,000 PCFs, there are 5 \emph{billion} integrals. 

We see a growing need to work with datasets of this size (and beyond). This is especially true within biology, where single-cell sequencing techniques are now producing samples with observations in the millions \citep{jovic2022single}. Combining these types of datasets with subsampling-based TDA methods (e.g., \cite{agerberg2023global}) is fertile ground for large-scale exploration using the tools in this article.

On the other hand, there is value in also being fast for smaller computations since this enables construction of, e.g., efficient iterative schemes. Likewise, for hyperparameter tuning, faster computation means that a larger hyperparameter space can be explored.


Central to our approach is a procedure we call \emph{rectangle iteration}. This will be used both in computing pairwise integrals, as well as averages, standard deviations, and so on. For transformations and computations involving a single PCF, rectangle iteration devolves into a simpler scheme called \emph{segment iteration}.

We note that while \citet{cohen1999finding} presents a similar scheme to ours for computing integrals, our methodology is more general in that it is not restricted to computing a specific integral (Earth Mover's Distance) and that it is applicable also for computing combinations of PCFs.

Our main contribution is an easy-to-use GPU-accelerated package that implements the aforementioned procedures and serves as a general toolbox for working with PCFs. We envision that the software will be used both as a standalone package, as well as a building block in a larger context on the level of performance primitives. For users without a CUDA-enabled GPU, we provide a parallel CPU-based implementation.

Existing approaches and software for doing computations with PCFs that we are aware of rely on constructing a common grid in-place for each computation \citep{bubenik2017persistence, dlotko2022euler}, breaking up the input functions along a fixed grid \citep{giotto-tda}, or fitting of a smooth function to each PCF \citep{lu2008reproducing}. The first of these results in computations at machine precision, whereas the others do not. Usually, there is a specific curve in mind (e.g., the Euler characteristic curve), rather than our more general approach as a tool for any piecewise invariant. None run on GPU. 

We note that while using a fixed grid certainly leads to fast computations, this approach requires the user to select a grid up front. One way to avoid this is to construct a grid using all the points at which any of the functions of interest change. This leads to simple computations but at the cost of using a potentially huge amount of memory. Also, any speed advantage that is gained by using common grids is likely lost due to the sheer amount of grid points required.

Our approach is different in that it computes a common grid implicitly for each computation. Beyond the obvious benefit of not having to choose a grid, integrals can be computed without any costly memory reallocations and the algorithm is asymptotically optimal, running in linear time in the number of time points in the PCFs.

We also considered packages that compute distances between time series. For \proglang{R}, there is the \pkg{proxy} package \citep{proxy}, but this package treats a pair of time series as two equal-length lists of numbers. This is not directly applicable to our case as the type of time series we consider are typically neither uniformly spaced nor of equal length. We should also mention the \pkg{TSDist} package \citep{tsdist}. However, this package uses \pkg{proxy} internally.

Finally, one may object that it is unnecessary to work with PCFs at ``full fidelity'' and that approximations are sufficient for most practical purposes. We take the view that this is an upstream problem. Given some representation of the data in terms of PCFs, be they exact up to machine precision, or some approximated version, the computations should still be as fast as possible. In other words, we view our framework as an enabling technology for working with PCFs efficiently.

The paper is structured as follows. In Section \ref{sec:pcfs}, we give the necessary background on PCFs and their reductions/combinations. In Section \ref{sec:integration} we present algorithms for computing integrals of combinations of PCFs. This is followed up in Section \ref{sec:reductiontrees}, where we build on the previous section to compute reductions of PCFs, including averages. Then, we give a short primer on how to use the \proglang{Python} module in Section \ref{sec:userguide}. This is followed, in Section \ref{sec:implementation}, by details on the implementation and some performance benchmarks. Finally, in Section \ref{sec:future}, we give some ideas for future directions.

\section[Piecewise constant functions]{Piecewise constant functions} \label{sec:pcfs}

We say that $f \colon [0,\infty) \to \mathbb{R}$ is a \emph{piecewise constant function} (PCF) if there is an ordered finite sequence of numbers $0=t_0 < t_1 < \cdots < t_n$ such that $f$ is constant on each of the intervals $[t_0,t_1)$, $[t_1,t_2), \ldots, [t_n,\infty)$. The sequence $t_{0:n}$ is said to \emph{discretize} $f$, and we write $(f,t_{0:n})$ to denote the function-sequence pair. We think of a PCF as a time series of numbers.

If $t'_{0:n'}$ is another ordered finite sequence containing the elements of $t_{0:n}$, we say that $t'_{0:n'}$ is a refinement of $t_{0:n}$, and we write $t_{0:n} \leq t'_{0:n'}$. One sees that this forms a partial order on the class of discretizations. Clearly, if $t_{0:n} \leq t'_{0:n'}$ and $t_{0:n}$ discretizes $f$, then $t'_{0:n'}$ also discretizes $f$. Moreover, since $f$ changes only at finitely many points, there is a unique \emph{minimal discretization} of $f$.

If we have two PCFs, $(f,t_{0:n}), (g,t'_{0:n'})$, discretized by $t_{0:n}$ and $t'_{0:n'}$, respectively, a \emph{common discretization} $s_{0:m}$ is an ordered sequence of time points with both $t_{0:n} \leq s_{0:m}$ and $t'_{0:n'} \leq s_{0:m}$. If $s_{0:m} \leq s'_{0:m'}$ for any common discretization $s'_{0:m'}$, we say that $s_{0:m}$ is a \emph{minimal common refinement} of $t_{0:n}$ and $t'_{0:n'}$.

A discretization is a finite subposet of the poset $[0, \infty)$. A collection of such subposets is a finite-type distributive lattice with the union as the join and the intersection as the meet. This structure guarantees that minimal common refinements exist.

A PCF $(f,t_{0:n})$ can be stored efficiently in computer memory in the form of a $n \times 2$ matrix
\begin{equation*}
    [t_{0:n};f] \colonequals \begin{bmatrix}
        0 & t_1 & \cdots & t_n \\
        f(0) & f(t_1) & \cdots & f(t_n)
    \end{bmatrix}^T.
\end{equation*}
Occasionally, we will write $t(A)_i$ and $v(A)_i$ to refer to the time and value, respectively, in the $i$-th row (starting from $0$) of a PCF matrix $A$.

Next, we discuss operations on PCFs. We define addition and scalar multiplication of PCFs in the usual pointwise sense for functions. That is, if $a \in \mathbb{R}$ and $f$ and $g$ are PCFs, then $(af)(t) \colonequals af(t)$ and $(f+g)(t) \colonequals f(t)+g(t)$ for all $t \geq 0$. The zero PCF, $0$, is the PCF that is identically zero, and similarly for the one PCF, $1$. We see that PCFs form a real vector space, which we will denote by $\Pcf$.

By restricting our attention to subsets of $\Pcf$, we can obtain a variety of constructions that are useful in analysis. Such subsets may or may not be vector subspaces of $\Pcf$. Common choices of subsets are PCFs that are nonnegative, nonincreasing, do not take the value $0$, and so on. None of these are vector subspaces. On the other hand, PCFs that are \emph{eventually $0$}, meaning that there exists some $T > 0$ such that $f(t)=0$ for all $t > T$, form a vector subspace.

The range of a PCF, $f$, is the set of all values attained by $f$ over its domain and is denoted by $\range(f)$. For a subset $\mathcal{F} \subseteq \Pcf$, we define $\range(\mathcal{F})$ to be the set $\bigcup_{f \in \mathcal{F}} \range(f)$.

Suppose $\mathcal{F}$ and $\tilde{\mathcal{F}}$ are two (possibly the same) subsets of $\Pcf$. A mapping $T \colon \mathcal{F} \to \tilde{\mathcal{F}}$ is called a \emph{transformation}. If $T(af+g)=aT(f)+T(g)$ for all $a \in \mathbb{R}$ and $f, g \in \mathcal{F}$, we say that $T$ is a \emph{linear} transformation.

Any function $h \colon X \to \mathbb{R}$ whose domain $X \subseteq \mathbb{R}$ contains $\range(\mathcal{F})$ is said to be \emph{admissible} for $\mathcal{F}$. Such a $h$ induces a transformation $h_* \colon \mathcal{F} \to h\mathcal{F}$ defined by $h_*(f)(t) \colonequals h(f(t))$, where $h\mathcal{F} = \{h \circ f : f \in \mathcal{F}\}$ is called the \emph{postcomposition} of $\mathcal{F}$ by $h$. Note that $h\mathcal{F}$ is itself a subset of $\Pcf$. Expressions such as $|f|$ and $f^2$ are to be understood to mean that we apply the transformations induced by $h(x)=|x|$ and $h(x)=x^2$, respectively, and so on.

\subsection[Combinations and reductions]{Combinations and reductions}

Often, we are interested in working with combinations of multiple PCFs. Let $\mathcal{F}_1$ and $\mathcal{F}_2$ be subsets of $\Pcf$. A binary operation $\bowtie \colon \mathcal{F}_1 \times \mathcal{F}_2 \to \Pcf$ is called a \emph{combination operation}. Although this definition is quite weak, it will lead to some useful constructions, as we will see shortly. A stronger notion is that of a \emph{reduction operation} on a subset $\mathcal{F} \subseteq \Pcf$, by which we mean a binary operation $\oplus \colon \mathcal{F} \times \mathcal{F} \to \mathcal{F}$ that is both associative and commutative.

Since reduction operations can be applied in any order, expressions such as $\bigoplus_{f_i \in \mathcal{C}} f_i$ for some finite collection $\mathcal{C}\subset \mathcal{F}$ of members of $\mathcal{F}$ are well-defined. We call an expression of this form a \emph{reduction}. It is a map $\mathcal{F}^n \to \mathcal{F}$ for some $n \in \mathbb{N}$. 

In the same vein, an expression in terms of combination operations and PCFs is called a \emph{combination}. Since the vector space operations of $\Pcf$ can be viewed as combinations, this naturally leads to \emph{linear combinations}, which are defined as combinations that are consist solely of vector space operations.

Both combination and reduction operations can be easily constructed from more familiar functions. Let $\mathcal{F}_1, \mathcal{F}_2 \subseteq \Pcf$,
and $X_1, X_2 \subseteq \mathbb{R}$, and let $h \colon X_1 \times X_2 \to \mathbb{R}$ be any function. If $\range(\mathcal{F}_i) \subseteq X_i$ for $i=1,2$, we say that $h$ is \emph{admissible} for $\mathcal{F}_1 \times \mathcal{F}_2$. In the case that $\mathcal{F}_1 = \mathcal{F}_2$, we say simply that $h$ is admissible for $\mathcal{F}_1$.

An admissible function, $h$, induces a combination operation $h_* \colon \mathcal{F}_1 \times \mathcal{F}_2 \to \Pcf$ defined by $h_*(f_1,f_2)(t) \colonequals h(f_1(t),f_2(t))$ for $f_1 \in \mathcal{F}_1$ and $f_2 \in \mathcal{F}_2$. Moreover, if $\mathcal{F}_1 = \mathcal{F}_2$ and if $h$ is associative and symmetric, the induced operation is a reduction operation. In particular, any function $\mathbb{R}^2 \to \mathbb{R}$ is admissible for $\Pcf$. Likewise, if $\mathcal{F}$ consists of PCFs that take only nonnegative values, any function $h \colon [0, \infty)^2 \to \mathbb{R}$ is admissible for $\mathcal{F}$. 

Functions that induce combination and reduction operations in this manner will be called \emph{combination functions} and \emph{reduction functions}, respectively. It is to be understood from the context that a function labeled in this way is assumed to be admissible for the particular subsets of interest.

As a concrete example of the preceding construction, consider the reduction operation $\oplus=h_*$ induced by the map $h(a,b)=a+b$. The average of a collection $\mathcal{C}$ of $n$ PCFs, $f_1,f_2,\ldots,f_n$, can then be computed as $\overline{f}=\frac{1}{n}\bigoplus_{i=1}^n f_i$.

The application of a combination operation with a fixed right-hand (or left-hand) side induces a transformation. As usual, we let $\mathcal{F}$ and $\tilde{\mathcal{F}}$ be subsets of $\Pcf$, and then we let $\bowtie \colon \mathcal{F} \times \tilde{\mathcal{F}} \to \Pcf$ be a combination operation. For an $\tilde{f} \in \tilde{\mathcal{F}}$, we can then define a transformation $T \colon \mathcal{F} \to \Pcf$ by $T(\bullet) \colonequals \bullet \bowtie \tilde{f}$. In the case of an induced combination $h_*$, the corresponding transformation is $h_*(\bullet, \tilde{f})$.

From this, we can compute the (unbiased) sample standard deviation $\hat{\sigma}(\mathcal{C})$ of the collection $\mathcal{C}$. Using $\overline{f}$ that we computed earlier, we let $h(a,b)=(a-b)^2$ and consider the transformation $h_*(\bullet, \overline{f})$. Then, $\hat{\sigma}(\mathcal{C}) = \sqrt{\frac{1}{n+1} \bigoplus_{i=1}^n h_*(f_i,\overline{f})}$, again with $\oplus$ induced by the map $(a,b) \mapsto a+b$. We note that although $h$ is only a combination function (from the lack of associativity), the expression $\hat{\sigma}(\mathcal{C})$ is a reduction.

One may wish to weight certain parts of a combination or reduction. We might, for example, have some intuition for why certain regions of a domain are more important, or we might try to learn such a dependence. For this purpose, we consider three-parameter versions of combination/reduction functions, $h \colon X_1 \times X_2 \times [0, \infty) \to \mathbb{R}$, for some $X_1,X_2 \subseteq \mathbb{R}$. 

As usual, if $X_1 \supseteq \range(\mathcal{F}_1)$ and $X_2 \supseteq \range(\mathcal{F}_2)$ for some $\mathcal{F}_1,\mathcal{F}_2 \subseteq \Pcf$, we say that $h$ is admissible for $\mathcal{F}_1 \times \mathcal{F}_2$, and we define $h_* \colon \mathcal{F}_1 \times \mathcal{F}_2 \to \Pcf$ by $h_*(f_1,f_2)(t) \colonequals h(f_1(t),f_2(t),t)$ to be the \emph{time-dependent combination operation induced by} $h$. 

Suppose now that we have a map of the form $h \colon X^2 \times [0, \infty) \to X$, for some $X \subseteq \mathbb{R}$, that is admissible for $\mathcal{F} \times \mathcal{F} \subseteq \Pcf \times \Pcf$. If at each fixed $t_0$, the map $h(-,-,t_0) \colon X^2 \to X$ is associative and symmetric, we say that $h_*$ is a \emph{time-dependent induced reduction}. 

\subsection{Functionals}

A functional on a subset $\mathcal{F} \subseteq \Pcf$ is a mapping $\mathcal{F} \to \mathbb{R}$. Similarly, for a product space $\mathcal{F} \times \tilde{\mathcal{F}} \subseteq \Pcf \times \Pcf$, a functional is a mapping $\mathcal{F} \times \tilde{\mathcal{F}} \to \mathbb{R}$. For statistical analysis and optimization, etc., we are often interested in the value of a functional when applied to a given PCF or pair of PCFs. In the single-PCF case, we may, for example, be interested in the functional $\max_{t} f(t)$ acting on the PCF $f$. Such functionals are easily computed by simple iteration over $[t_{0:n};f]$, or by using a parallel reduction scheme.

The two-PCF case is more interesting. Let $\mathcal{F} \subseteq \Pcf$ and let $h_*$ be an induced combination from an $\mathcal{F} \times \mathcal{F}$-admissible function $h$. We are primarily interested in functionals of the form
\begin{align*}
    F(f,g) = r\left( \int_a^b h_*(f,g) \, dt \right)
\end{align*}
where $0 \leq a < b \leq \infty$ and $r \colon X \to \mathbb{R}$ is a function whose domain, $X$, contains the set $\{\int_a^b h_*(f,g) \, dt : f,g \in \mathcal{F} \} $.  We will refer to a functional of this form as a \emph{combination integral}. The result of applying a combination integral on two PCFs, $f_1$ and $f_2$, is an \emph{integrated combination} of $f_1$ and $f_2$. Furthermore, we say that $F$ is symmetric if $h$ is symmetric.

Within this framework, we can, for example, recover the $L_2$ inner product by letting $h(a,b)=ab$ and taking $r(x)=x$. Similarly, the $L_p$ distance has $h(a,b)=|a-b|^p$ and $r(x)=x^{1/p}$. 

By using time-dependent combinations, more elaborate constructions are possible, as well. If $w \colon [0,\infty) \to [0,\infty)$ is an integrable function, we can compute a weighted $L_p$ distance as $\left(\int_0^\infty |f_1(t)f_2(t)|^p w(t) \, dt\right)^{1/p}$ by letting $r(x)=x^{1/p}$ and using $h(a,b,t)=|a-b|^pw(t)$. We can, of course, follow a similar recipe to obtain a weighted $L_2$ inner product.

\subsection{Generalized induced maps}

We note that the choice of having combination and reduction functions take values in $\mathbb{R}$ was necessary for the induced maps to produce PCFs. However, we could consider maps $\mathcal{F}_1 \times \mathcal{F}_2 \to \mathcal{M}$ for $\mathcal{F}_1, \mathcal{F}_2 \subset \Pcf$ into a more general space $\mathcal{M}$, such as a manifold or an arbitrary topological or function space.

Additionally, if $h \colon X \to \mathcal{M}$ is admissible for $\mathcal{F} \subset \Pcf$ (again meaning that $\range(f) \subset X \subset \mathbb{R}$), the induced map $h_* \colon \mathcal{F} \to \mathcal{M}$ gives fixed points on $\mathcal{M}$ over discrete time intervals. One could then consider the family of sets $\mathcal{A}_t(f) \colonequals \{h_*(f)(s) : s \leq t\}$ for $f \in \mathcal{F}$, giving a \emph{filtration} of the space $\mathcal{A}_\infty(f)$.

If, for example, $\mathcal{M}$ is a Riemannian manifold, there is a well-defined distance between points, and we can view $\mathcal{A}_\infty(f)$ a filtered distance space on which we can perform further topological analysis.

For now, we will leave this construction as a remark of something that could be interesting to explore in a future work. In the next section, we will return to the more concrete problem of integrating PCFs that is the main topic of this paper.

\section{Integration of PCF combinations} \label{sec:integration}

Let $\mathcal{F} \subseteq \Pcf$ and take $f, g \in \mathcal{F}$. Let $h$ be a (time-dependent) combination function that is admissible for $\mathcal{F}$. We would like to compute the integral
\begin{align}
\label{eq:integral}
    I \colonequals \int_a^b h_*(f,g)(t) \, dt.
\end{align}
If we have access to a common discretization, $s_{0:m}$, of $f$ and $g$, the integral above can be written as
\begin{align*}
    I = \sum_{i=0}^{m-1} \int_{s_i}^{s_{i+1}} h(f(s_i),g(s_i),t) \, dt,
\end{align*}
and in the time-independent combination case, the expression becomes even simpler:
\begin{align}
\label{eq:simpleint}
    I = \sum_{i=0}^{m-1} h(f(s_i),g(s_i)) (s_{i+1}-s_i).
\end{align}
\tikzset{pics/integration scan/.style = {
    code={

    \tikzstyle{a}=[color=blue, thick];
    \tikzstyle{b}=[color=red, thick];
    \tikzstyle{aa}=[arrow, thick, blue];
    \tikzstyle{ba}=[arrow, thick, red];

    \tikzstyle{s}=[black, dashed];

    \def\m{0.7}

    \foreach \i in {0,...,6}{
        \draw ({\m*\i}, -0.5) node {\i};
    }

    \foreach \i in {0,...,4}{
        \draw (-0.5, {\m*\i}) node {\i};
    }

    \foreach \i in {0,...,3}{
        \coordinate (tbluebox\i) at ({0.5+0.5*\i}, -2);
        \coordinate (vbluebox\i) at ({0.5+0.5*\i }, -2.5);

        \draw (vbluebox\i) rectangle +(0.5, 0.5);
        \draw (tbluebox\i) rectangle +(0.5, 0.5);
    }

    \foreach \i in {0,...,1}{
        \coordinate (tredbox\i) at ({3.5+0.5*\i}, -2);
        \coordinate (vredbox\i) at ({3.5+0.5*\i }, -2.5);
    
        \draw (vredbox\i) rectangle +(0.5, 0.5);
        \draw (tredbox\i) rectangle +(0.5, 0.5);
    }

    \node[color=blue] at ($(tbluebox0) + (0.25,0.25)$) {0};
    \node[color=blue] at ($(tbluebox1) + (0.25,0.25)$) {2};
    \node[color=blue] at ($(tbluebox2) + (0.25,0.25)$) {3};
    \node[color=blue] at ($(tbluebox3) + (0.25,0.25)$) {5};

    \node[color=blue] at ($(vbluebox0) + (0.25,0.25)$) {4};
    \node[color=blue] at ($(vbluebox1) + (0.25,0.25)$) {3};
    \node[color=blue] at ($(vbluebox2) + (0.25,0.25)$) {1};
    \node[color=blue] at ($(vbluebox3) + (0.25,0.25)$) {0};

    \node[color=red] at ($(tredbox0) + (0.25,0.25)$) {0};
    \node[color=red] at ($(tredbox1) + (0.25,0.25)$) {4};

    \node[color=red] at ($(vredbox0) + (0.25,0.25)$) {2};
    \node[color=red] at ($(vredbox1) + (0.25,0.25)$) {0};

    \ifnum #1 > 0
        \fill[gray!20] (0, {\m*2}) rectangle ({\m*2}, {\m*4});
        \draw (0, {\m*2}) rectangle ({\m*2}, {\m*4});
    \fi
    
    \ifnum #1 > 1
        \fill[gray!20] ({\m*2}, {\m*2}) rectangle ({\m*3}, {\m*3});
        \draw ({\m*2}, {\m*2}) rectangle ({\m*3}, {\m*3});
    \fi

    \ifnum #1 > 2
        \fill[gray!20] ({\m*3}, {\m*1}) rectangle ({\m*4}, {\m*2});
        \draw ({\m*3}, {\m*1}) rectangle ({\m*4}, {\m*2});
    \fi

    \ifnum #1 > 3
        \fill[gray!20] ({\m*4}, 0) rectangle ({\m*5}, {\m*1});
        \draw ({\m*4}, 0) rectangle ({\m*5}, {\m*1});
    \fi
    
    \draw [a] (0,{\m*4}) -- ({\m*2},{\m*4});
    \draw [a] ({\m*2},{\m*4}) -- ({\m*2},{\m*3});
    \draw [a] ({\m*2},{\m*3}) -- ({\m*3},{\m*3});
    \draw [a] ({\m*3},{\m*3}) -- ({\m*3},{\m*1});
    \draw [a] ({\m*3},{\m*1}) -- ({\m*5},{\m*1});
    \draw [a] ({\m*5},{\m*1}) -- ({\m*5},{\m*0});
    \draw [a] ({\m*5},0) -- ({\m*6},0);

    \draw [b] (0, {\m*2}) -- ({\m*4}, {\m*2});
    \draw [b] ({\m*4}, {\m*2}) -- ({\m*4}, 0);
    \draw [b] ({\m*4}, 0) -- ({\m*6}, 0);

    \ifnum #1 = 0
        \draw [aa] (0, {\m*5}) -- (0, {\m*4.5}); 
        \draw [aa] ($(vbluebox0)+(0.25,-0.45)$) -- ($(vbluebox0)+(0.25,-0.125)$);
    \fi
    \ifnum #1 = 1
        \draw [aa] ({\m*2}, {\m*5.75}) -- ({\m*2}, {\m*5});
        \draw [aa] ($(vbluebox1)+(0.25,-0.45)$) -- ($(vbluebox1)+(0.25,-0.125)$);
    \fi
    \ifnum #1 = 2
        \draw [aa] ({\m*3}, {\m*5.75}) -- ({\m*3}, {\m*5});
        \draw [aa] ($(vbluebox2)+(0.25,-0.45)$) -- ($(vbluebox2)+(0.25,-0.125)$);
    \fi
    \ifnum #1 = 3
        \draw [aa] ({\m*3}, {\m*5.75}) -- ({\m*3}, {\m*5});
        \draw [aa] ($(vbluebox2)+(0.25,-0.45)$) -- ($(vbluebox2)+(0.25,-0.125)$);
    \fi
    \ifnum #1 = 4
        \draw [aa] ({\m*5}, {\m*5.75}) -- ({\m*5}, {\m*5});
        \draw [aa] ($(vbluebox3)+(0.25,-0.45)$) -- ($(vbluebox3)+(0.25,-0.125)$);
    \fi

    \ifnum #1 < 3
        \draw [ba] (0, -1.25) -- (0, -0.75); 
        \draw [ba] ($(vredbox0)+(0.25,-0.45)$) -- ($(vredbox0)+(0.25,-0.125)$);
    \fi
    \ifnum #1 > 2
        \draw [ba] ({\m*4}, -1.25) -- ({\m*4}, -0.75);
        \draw [ba] ($(vredbox1)+(0.25,-0.45)$) -- ($(vredbox1)+(0.25,-0.125)$);
    \fi

    \draw[arrow] (0, -0.1) -- (0, {\m*5.0}) node[above]{$v$};
    \draw[arrow] (-0.1, 0) -- ({\m*6.5}, 0) node[right]{$t$};

    \draw[red,fill=red] (0, {\m*2}) circle (.075);
    \draw[red,fill=white] ({\m*4}, {\m*2}) circle (.075);
    \draw[red,fill=red] ({\m*4}, 0) circle (.075);

    \draw[blue,fill=blue] (0, {\m*4}) circle (.075);
    \draw[blue,fill=white] ({\m*2}, {\m*4}) circle (.075);
    \draw[blue,fill=blue] ({\m*2}, {\m*3}) circle (.075);
    \draw[blue,fill=white] ({\m*3}, {\m*3}) circle (.075);
    \draw[blue,fill=blue] ({\m*3}, {\m*1}) circle (.075);
    \draw[blue,fill=white] ({\m*5}, {\m*1}) circle (.075);
    \draw[blue,fill=blue] ({\m*5}, 0) circle (.075);
    
    \ifnum #1 = 0
        \draw [s] (0, -0.25) -- (0, {\m*4.5});
    \fi
    \ifnum #1 = 1
        \draw [s] ({\m*2}, -0.25) -- ({\m*2}, {\m*5});
    \fi
    \ifnum #1 = 2
        \draw [s] ({\m*3}, -0.25) -- ({\m*3}, {\m*5}); 
    \fi
    \ifnum #1 = 3
        \draw [s] ({\m*4}, -0.25) -- ({\m*4}, {\m*5});
    \fi
    \ifnum #1 = 4
        \draw [s] ({\m*5}, -0.25) -- ({\m*5}, {\m*5}); 
    \fi

    \draw ($(vbluebox0) + (-0.3,0.25)$) node {$v_f$};
    \draw ($(tbluebox0) + (-0.25,0.25)$) node {$t$};

    \draw ($(vredbox0) + (-0.3,0.25)$) node {$v_{g}$};
    \draw ($(tredbox0) + (-0.25,0.25)$) node {$t$};

}}}

\begin{figure}
     \centering
     \begin{subfigure}[b]{0.3\textwidth}
         \centering
            \begin{tikzpicture}[scale=0.82, every node/.style={scale=0.82}]
                \pic at (0,0) {integration scan=1};
            \end{tikzpicture}
     \end{subfigure}
     \hfill
     \begin{subfigure}[b]{0.3\textwidth}
         \centering
            \begin{tikzpicture}[scale=0.82, every node/.style={scale=0.82}]
                \pic at (0,0) {integration scan=2};
            \end{tikzpicture}
     \end{subfigure}
     \hfill
     \begin{subfigure}[b]{0.3\textwidth}
         \centering
            \begin{tikzpicture}[scale=0.82, every node/.style={scale=0.82}]
                \pic at (0,0) {integration scan=3};
            \end{tikzpicture}
     \end{subfigure}
        \caption{Three steps of rectangle iteration. In the first step, the pointer belonging to $f$ (function drawn in blue) moves since a change in $f$ occurs before $g$. This is also true for the second step, but in the third step, $g$ now changes before $f$, so the second pointer moves. The scanline, shown as a dashed line, moves together with the last pointer shift and determines the current time point. The (transposed) matrix representations of $f$ and $g$ are shown below the plot, together with the respective pointers.}
        \label{fig:rectiter}
\end{figure}
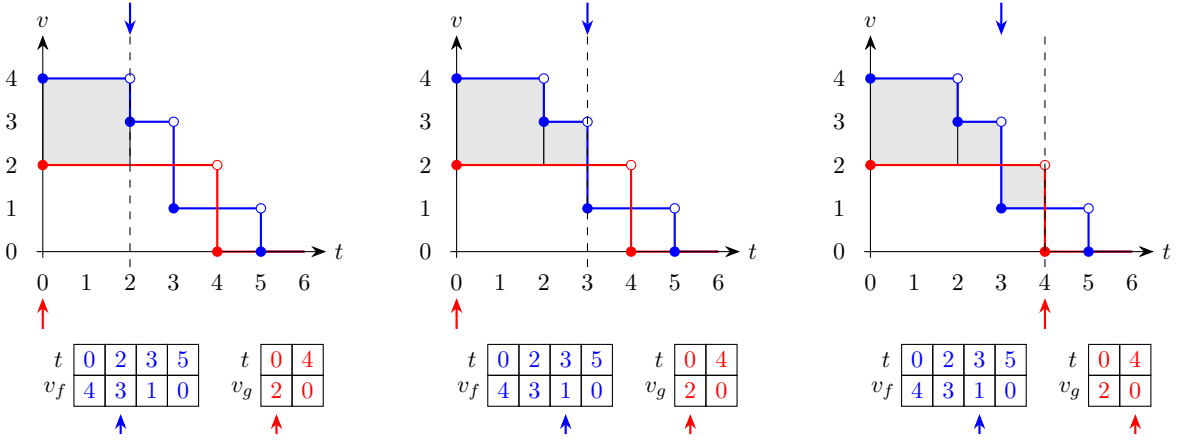
This motivates a procedure we call \emph{rectangle iteration} and which is summarized in Algorithm \ref{alg:iteraterects}. This computational scheme has the advantage that it does not require us to precompute a common discretization but rather one is computed implicitly during the evolution of the algorithm. Because of this, from an implementation standpoint, we can perform the computation of $I$ without allocating any additional memory on the heap. This, in conjunction with storing the PCFs contiguously in memory should lead to a low rate of cache misses, which is crucial for high performance on modern hardware.

By a \emph{rectangle}, we mean a four-tuple $(l,r,v_f,v_g) \in [0,\infty]^2 \times \mathbb{R}^2$, where the first two coordinates are time points (left and right) and the last two coordinates are function values of $f$ and $g$, respectively. Our idea is then to walk from left to right over the common discretization $s_{0:m}$, recording the function values at the left time point. Equation \eqref{eq:simpleint} can then be computed by simply summing up the contributions $(r-l)h(v_f,v_g)$ of each rectangle (see Algorithm \ref{alg:riemanntimeindep}).
\begin{algorithm}[t!]
\DontPrintSemicolon
\caption{\code{IterateRectangles}($[t_{0:n};f]$, $[t'_{0:n'};g]$, $a$, $b$)}
\label{alg:iteraterects}
\KwIn{PCFs $[t_{0:n};f]$ and $[t'_{0:n'};g]$, integration bounds $0 \leq a < b \leq \infty$}
\KwOut{List of rectangles $\mathcal{L}$ ordered by time of occurrence}
    $\mathcal{L} \gets $ empty list\;
    $t \gets a$\;
    $k \gets \max\{i : t_i \leq a\}$\;
    $k' \gets \max\{i : t'_i \leq a\}$\;
    \While{$t < b$}{
        $t_{\mathrm{prev}} \gets t$\;
        $v_f \gets f(t_k)$\;
        $v_g \gets g(t'_{k'})$\;
        \eIf{$k+1<n$ and $k'+1<n'$}{
            $\delta \gets (t_{k+1} - t'_{k'+1})$\;
            \If{$\delta \leq 0$}{
                $k \gets k + 1$\;
            }
            \If{$\delta \geq 0$}{
                $k' \gets k'+1$\;
            }
        }{
            \uIf{$k+1 < n$} {
                $k \gets k + 1$\;
            } \uElseIf{$k'+1 < n'$}{
                $k' \gets k' + 1$\;
            } \Else {
                Append rectangle with $l=t_{\mathrm{prev}}$, $r=b$, $v_f=v_f$, $v_g=v_g$ to $\mathcal{L}$\;
                \Return\;
            }
        }
        $t \gets \min\{\max\{t_k,t'_{k'}\}, b \}$\;
        Append rectangle with $l=t_{\mathrm{prev}}$, $r=t$, $v_f=v_f$, $v_g=v_g$ to $\mathcal{L}$\;
    }
\end{algorithm}

As alluded to earlier, one does not need to precompute a common discretization. Algorithm \ref{alg:iteraterects} works by keeping a pair of pointers into $[f,t_{0:n}]$ and $[g,t'_{0:n'}]$, respectively, starting from the left. At each step of the algorithm, we look one time point ahead for each of the PCFs. If $f$ changes before $g$, the second pointer moves ahead. If $g$ changes before $f$, the first pointer moves. And if $f$ and $g$ both change at the same time, then both pointers are moved ahead simultaneously. Readers that are familiar with computational geometry may recognize this as a scanline algorithm, only that we do not need to do any sorting up front since the PCFs are already ordered. The procedure is illustrated graphically in Figure \ref{fig:rectiter}

One does not need to store the list of rectangles (which would incur unnecessary heap allocations). Rather, we trigger a callback on each rectangle. For an efficient implementation, this can be done without incurring the cost of a function call at each step by relying on the inlining capabilities of an optimizing compiler, or by hand coding for a specific use case.

From these observations, we note rectangle iteration has linear runtime in the size of the largest PCF (in terms of the number of time points).
\begin{algorithm}[t!]
\DontPrintSemicolon
\caption{\code{CombineIntegrate}($[t_{0:n};f]$, $[t'_{0:n'};g]$, $a$, $b$, $h(x,y)$)}
\label{alg:riemanntimeindep}
\KwIn{$[t_{0:n};f]$, $[t'_{0:n'};g]$, $a$, $b$ as in Algorithm \ref{alg:iteraterects}; admissible combination function $h(x,y)$}
\KwOut{$I = \int_a^b h_*(f,g)(t) \, dt=\int_a^b h(f(t),g(t)) \, dt$}
$I \gets 0$\;
$\mathcal{L} \gets$ \code{IterateRectangles}($[t_{0:n};f]$, $[t'_{0:n'};g]$, $a$, $b$)\;
\ForEach{Rectangle $R$ in $\mathcal{L}$} {
    $\Delta t \gets (R.r - R.l)$\;
    $I \gets I +  h(R.v_f, R.v_g)\Delta t$\;
}
\end{algorithm}

Now suppose that $h$ is a time-dependent combination function with known antiderivative $H$ in $t$. Then, since $f$ and $g$ are constant over $(s_i, s_{i+1})$, $i=0,\ldots,m-1$, Equation \eqref{eq:integral} can be computed as
\begin{align*}
    I &= \sum_{i=0}^{m-1} [ H(f(s_i),f'(s_i),s_{i+1}) - H(f(s_i),f'(s_i), s_i)],
\end{align*}
by the fundamental theorem of calculus. The only change necessary in Algorithm \ref{alg:riemanntimeindep} is that the contribution for each rectangle changes from $h(R.v_f,R.v_g)\Delta t$ to $H(R.v_f,R.v_g,R.r) - H(R.v_f,R.v_g,R.l)$.

As an aside, it may be of interest to note that if $H$ is a parameterized by $\theta \in \Theta$, where $\Theta$ is a subset of $\mathbb{R}^d$ for some $d$, then $\partial I/\partial \theta_j$ can be calculated analytically as long as this can be done for $H$. We foresee that this could lead to an approach where $H$ is a neural network and we use an optimization procedure to find an optimal $H$ (and thus $h$) for a specific task. We aim to explore this connection in a future work.

\subsection{Integrated combination matrices}

The input to many computational methods from statistics (and otherwise) is a matrix of pairwise similarity/difference ``scores'' between observations in a dataset. For example, in topological data analysis, a typical setting is that each observation is associated with a finite point cloud whose topology is represented via a PCF. We would then like to measure how ``close'' or ``far'' these observations are from one another by comparing the corresponding PCFs. In light of the discussion of previous sections, we view PCFs as observations, and the similarity/difference is then given by some integrated combination of pairs of PCFs.

Let $F$ be a combination integral, let $\mathcal{C}=\{f_1,\ldots,f_M\}$ be a collection of PCFs and denote by $F_{ij} \colonequals F(f_i,f_j)$, $i,j=1,\ldots,M$. We call the matrix
\begin{align*}
    F(\mathcal{C}) \colonequals \begin{bmatrix}
        F_{11} & F_{12} & \cdots & F_{1M} \\
        F_{21} & F_{22} & \cdots & F_{2M} \\
        \vdots & \vdots & \ddots & \vdots \\
        F_{M1} & F_{M2} & \cdots & F_{MM}
    \end{bmatrix}
\end{align*}
the \emph{pairwise integrated combination matrix} of $\mathcal{C}$ under $F$.

If $F$ is symmetric, $F(\mathcal{C})$ will be symmetric. Typically, this is the case we are interested in since both distances and inner products are symmetric (recalling that $\Pcf$ is a real vector space). One then usually uses standard names like ($L_p$) distance matrix for the former and Gram/inner product matrix for the latter.

\subsection{PCF integrals}

Having treated pairwise integrals, we return to the simpler problem of computing integrals involving a single PCF. Besides the obvious case of computing the integral of a PCF, we may also want to compute, for example, norms of PCFs.

Because the presentation will be very similar to the previous sections, we try to only state the necessary parts and work in slightly less detail than before.

By a segment we mean a triple of numbers, $(l,r,v)$, where $l \leq r$ are the left and right endpoints of a horizontal segment with vertical coordinate equal to $v$---the \emph{value}. We consider functionals, $F$, of the form
\begin{align*}
    F(f) = r\left(\int_a^b h(f(t)) \, dt\right),
\end{align*}
where $h$ is some admissible function, $0 \leq a < b \leq \infty$, and $r$ is a function whose domain is some suitable subset of $\mathbb{R}$ so that the expression above is well-defined.

The integral can computed using the procedure in Algorithm \ref{alg:iteratesegments}. The resulting list of segments from this algorithm are then considered one-by-one in turn, where for a segment $(l,r,v)$, the contribution to the integral is $(r-l)h(v)$. 

In other words, we replace rectangle iteration by \emph{segment iteration}, and we modify Algorithm \ref{alg:riemanntimeindep} accordingly. Time-dependent versions of the algorithms can be obtained easily by following a similar scheme as before.

\begin{algorithm}[t!]
\DontPrintSemicolon
\caption{\code{IterateSegments}($[t_{0:n};f]$, $a$, $b$)}
\label{alg:iteratesegments}
\KwIn{PCF $[t_{0:n};f]$, integration bounds $0 \leq a < b \leq \infty$}
\KwOut{List of segments $\mathcal{L}$, ordered by time of occurrence}
$\mathcal{L} \gets $ empty list\;
$t \gets a$\;
$k \gets \max\{i : t_i \leq a\}$\;
$v \gets f(t_k)$\;
\For{$i \in \{k+1,\ldots,n\}$}{
    Append segment with $l=t$, $r=\min\{t_i,b\}$, $v=v$ to $\mathcal{L}$\;
    \If{$t_i \geq b$}{
        \Return\;
    }
    $t \gets t_i$\;
    $v \gets f(t_i)$\;
}
Append segment with $l=t_n$, $r=b$, $v=v$ to $\mathcal{L}$ \;
\end{algorithm}

\section{Reductions} \label{sec:reductiontrees}

\tikzset{reduction/.style={draw,circle,append after command={
        [shorten >=\pgflinewidth, shorten <=\pgflinewidth,]
        (\tikzlastnode.north) edge (\tikzlastnode.south)
        (\tikzlastnode.east) edge (\tikzlastnode.west)
        }
    }
}

\begin{figure}
    \centering
    \begin{tikzpicture}[
    funcnode/.style={draw=black,  thick, minimum size=7mm}
    ]
        \foreach \i in {1,...,3}
        {
            \draw (\i, 10) node[funcnode] (f\i) {$f_\i$};
        }

        \foreach \i in {4,...,6}
        {
            \draw (0.5+\i, 10) node[funcnode] (f\i) {$f_\i$};
        }

        \foreach \i in {7,...,8}
        {
            \draw (2*0.5+\i, 10) node[funcnode] (f\i) {$f_{\i}$};
        }

        \draw(1, 8.5) node[reduction,scale=1.5,thick] (red0) {};
        \foreach \i in {1,...,3}
        {
            \draw (f\i.south) -- (red0);
        }

        \draw(0.5+4, 8.5) node[reduction,scale=1.5,thick] (red1) {};
        \foreach \i in {4,...,6}
        {
            \draw (f\i.south) -- (red1);
        }

        \draw(2*0.5+7, 8.5) node[reduction,scale=1.5,thick] (red2) {};
        \foreach \i in {7,...,8}
        {
            \draw (f\i.south) -- (red2);
        }

        \draw(1, 7.5) node[reduction,scale=1.5,thick] (red10) {};
        \draw (red0) -- (red10);
        \draw (red1) -- (red10);

        \draw(2*0.5+7, 7.5) node[reduction,scale=1.5,thick] (red11) {};
        \node[right of=red11] {No-Op};
        \draw (red2) -- (red11);

        \draw(1, 6.5) node[reduction,scale=1.5,thick] (red20) {};
        \draw (red10) -- (red20);
        \draw (red11) -- (red20);

    \end{tikzpicture}
    \caption{Example reduction tree on eight PCFs $f_1,\ldots,f_8$. Here, we split the PCFs in groups of at most three PCFs at the leaf level of the tree, although other configurations are possible. Typically, we will use binary reduction trees. One of the reductions (marked as No-Op) is inserted to make the programming a bit easier but could also be left out ($f_7 \oplus f_8$ would then have to feed directly into the last reduction on the left). Each column in the tree uses one reduction accumulator and at the end of the reduction we return the value stored in the leftmost column.}
    \label{fig:reductiontree}
\end{figure}
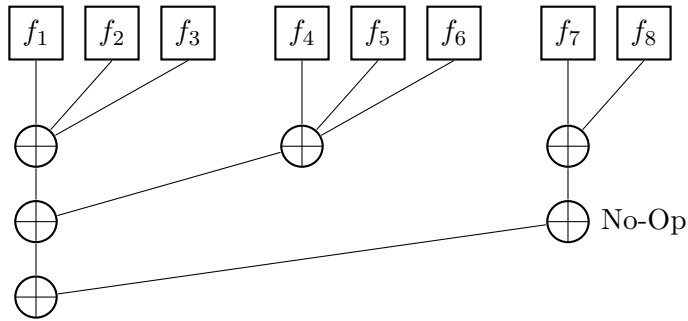

Since reduction operations are associative and commutative, they can be deterministically executed in arbitrary order.\footnote{Within machine precision, as usual, when using floating point numbers.} As such, we can execute reductions in a tree, like the one displayed in Figure \ref{fig:reductiontree}. Reduction trees are a standard pattern in parallel computing (see, e.g., \cite{hwu2022programming}). 

However, the usual case is that the reduced values are small in terms of memory footprint (e.g., a single number). In our case, the reduced object at each tree level is a PCF, which can potentially contain thousands of time points. For this reason, we would like to avoid costly memory reallocations as much as possible. In a na\"{i}ve configuration, each application of a reduction operation, $\oplus$, on two PCFs $f_1$ and $f_2$ allocates a new point buffer that stores the result of $f_1 \oplus f_2$. In a binary reduction tree over $M$ PCFs, $M-1$ applications of $\oplus$ are required \citep{hwu2022programming}, resulting in $M-1$ reallocations. The cost of growing the buffer generally increases at each tree level since the resulting PCF is at least as large as the biggest PCF on the previous level. Therefore, we would like to reuse memory as much as possible across the tree levels.

This can be done by introducing a \emph{reduction accumulator}. This is an instance, $A$, of an object $\mathcal{A}$ that contains an internal $\Pcf$ state, $A_p$, and supports three operations: $A \mathrel{\oplus}= \Pcf$, $A \mathrel{\oplus}= A'$, and $A \to \Pcf$, where $A'$ is a separate instance of $\mathcal{A}$. The first operation, $A \mathrel{\oplus}= \Pcf$ applies the reduction $\oplus$ on the internal state $A_p$ and the right-hand-side, and stores the result in $A_p$. The second operation, $A \mathrel{\oplus}= A'$ is the same as the first operation but with the right-hand-side equal to $A'_p$. Finally, $A \to \Pcf$ simply returns the internal state $A_p$.

In our implementation, we use a side buffer $\tilde{A}_p$ in $\mathcal{A}$ to compute the reduction and $A_p$ then gets swapped with $\tilde{A}_p$. This happens in $O(1)$ time.

\begin{algorithm}[t!]
\DontPrintSemicolon
\caption{\code{ReducePair}($[t_{0:n};f]$, $[t'_{0:n'};g]$, $h(x,y)$)}
\label{alg:rectreduce}
\KwIn{$[t_{0:n};f]$, $[t'_{0:n'};g]$ as in Algorithm \ref{alg:iteraterects}; combination function $h(x,y)$}
\KwOut{$h_*(f,g)$}
$H_* \gets  (n+n'+2) \times 2$ matrix\;
$i \gets 0$\;
$v_{\mathrm{last}} \gets 0$\;
\ForEach{Rectangle $R$ in $\mathcal{L}$} {
    $v \gets h(R.v_{f}, R.v_g)$\;
    $t \gets R.l$\;
    \If{$t = 0$ or $v \neq v_{\mathrm{last}}$}{
        $t(H_*)_i \gets t$\;
        $v(H_*)_i \gets v$\;
        $i \gets i + 1$\;
    }
    $v_{\mathrm{last}} \gets v$\;
}
$h_*(f,f') \gets $ PCF with matrix $H_*[0:i-1,:]$\;
\end{algorithm}

\subsection{Computing reduction pairs}

One issue remains. Until now, we have completely sidestepped the issue of computing the reduction $f \oplus g$ of two PCFs $f$ and $g$, discretized by, say $t_{0:n}$ and $t'_{0:n'}$, respectively. Here, we can again use rectangle iteration. We start by allocating enough memory to store a PCF whose size, $N$, is the sum of the two input PCF sizes. Then, we run rectangle iteration and, for each rectangle $\mathcal{R}$, we insert a time point $t$ at the left coordinate of $\mathcal{R}$ with value equal to the reduced value $h(f(t),g(t))$. We note that we only need to insert a new time point if the value differs from the last inserted time point, or if $t=0$. Since this procedure can result in at most a PCF of size $N$, the memory allocation from earlier is sufficient. We then truncate the PCF (i.e., by resizing the memory downwards) to match the actual size needed. The steps are summarized in Algorithm \ref{alg:rectreduce}.

Suppose we have $M$ PCFs to reduce, each of which has at most $n$ time points. Working from the top of the tree, there are $M/2$ applications of the reduction operation necessary, and each such operation runs in $O(n)$ time and produces a PCF of size at most $2n$. In the next step, there are $M/4$ applications of an $O(2n)$ reduction operation to produce a PCF of size at most $4n$, and so on. By induction, the $k$-th level, counting the first level as $k=1$, requires $M/2^k$ reduction operations of PCFs of size at most $2^kn$. There are $\log M$ levels in the tree, resulting in $\log M - 1$ reduction operations being necessary. Putting everything together, we get a total of
\begin{align*}
\sum_{k=1}^{\log M - 1} O(2^k n) \frac{M}{2^k} = \sum_{k=1}^{\log M - 1} O(nM)=O(nM\log M)
\end{align*}
operations necessary for the entire algorithm.

We should note that one could also perform the reduction by reducing the first pair of PCFs, then applying the reduction operation to this result together with the third PCF, and so on. This only requires $O(nM)$ operations but we lose the ability to parallelize. This approach could, however, be appealing if we have a large number of separate reductions as we could then parallelize on the level of reductions rather than individual reduction operations.

\section{User guide} \label{sec:userguide}

We provide the source code to our library and module in a GitHub repository\footnote{\url{https://github.com/bwehlin/masspcf}}. Users can also install the module from the Python Package Index (PyPI) via
\begin{CodeChunk}
\begin{CodeInput}
pip install masspcf
\end{CodeInput}
\end{CodeChunk}

After installing, we begin by importing the \proglang{Python} module. Typically, we would like to have access to \pkg{numpy}, as well.
\begin{CodeChunk}
\begin{CodeInput}
import masspcf as mpcf
import numpy as np
\end{CodeInput}
\end{CodeChunk}

We can then create a PCF from a $n \times 2$ array, where the first column represents time and the second column is the function value at each time.
\begin{CodeChunk}
\begin{CodeInput}
TV = np.array([[0, 4], [2, 3], [3, 1], [5, 0]])
f = mpcf.Pcf(TV)
\end{CodeInput}
\begin{CodeOutput}
<PCF size=4, dtype=float64>
\end{CodeOutput}
\end{CodeChunk}
As can be seen from the output, the created PCF will inherit its datatype from the \pkg{numpy} array. We note that it is often beneficial to use \code{np.float32} over \code{np.float64}, both for storing the result of large matrix integrations (due to the memory requirement), and for GPU users in particular, since 32-bit computations usually are much faster there than 64-bit.

For the reverse direction, to obtain the time-value matrix of a PCF, we support the \proglang{Python} buffer protocol\footnote{See \url{https://docs.python.org/3/c-api/buffer.html} for details.}. This makes it easy to store PCFs as, for example, \pkg{numpy} arrays.
\begin{CodeChunk}
\begin{CodeInput}
fMatrix = np.array(f)
\end{CodeInput}
\begin{CodeOutput}
[[0. 4.]
 [2. 3.]
 [3. 1.]
 [5. 0.]]
\end{CodeOutput}
\end{CodeChunk}
We note that the corresponding buffers are read-only as we must ensure that all PCFs start at zero and have their time points in increasing order, a contract that cannot be maintained if users are allowed to write directly into the PCF point buffer.

For plotting, we provide a convenience function \code{masspcf.plotting.plot(...)} that integrates with \pkg{matplotlib} \citep{hunter2007matplotlib}. Our plotting function calls \code{step(...)} from \code{pyplot} within \pkg{matplotlib} with \code{where='post'} on the points in the PCF.
\begin{CodeChunk}
\begin{CodeInput}
from masspcf.plotting import plot as plotpcf
import matplotlib.pyplot as plt

plotpcf(f)

plt.xlabel('t')
plt.ylabel('f(t)')
\end{CodeInput}
\end{CodeChunk}
\begin{figure}[H]
    \includegraphics[width=7cm]{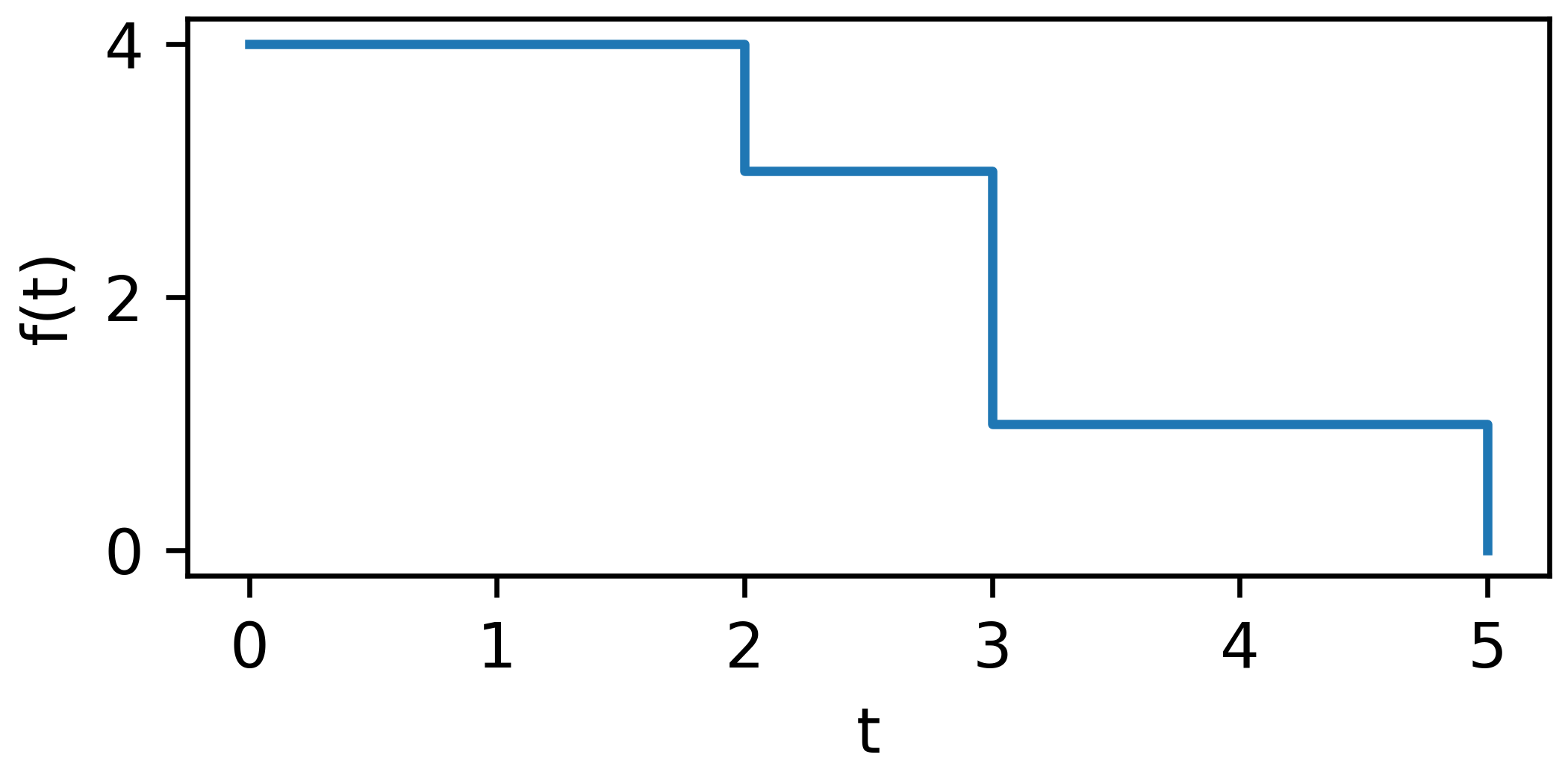}
\end{figure}

\subsection{Multidimensional arrays}

For most applications, however, it is preferable to work with multidimensional arrays of PCFs.

To declare, for example, a $10 \times 5 \times 4$ array, \code{Z}, of zero PCFs (i.e., PCFs that are constantly 0) and display its size, we can write
\begin{CodeChunk}
\begin{CodeInput}
Z = mpcf.zeros((10, 5, 4))
print(Z.shape)
\end{CodeInput}
\begin{CodeOutput}
Shape(10, 5, 4)
\end{CodeOutput}
\end{CodeChunk}

Views (lightweight objects that refer to arrays or parts of arrays without storing the underlying data) can be obtained using the usual \proglang{Python} array indexing syntax.
\begin{CodeChunk}
\begin{CodeInput}
Z[3, :, :]      # shape = (5,4), indices [3] x [0,...,4] x [0,...,3]
Z[2:9:3, 1:, 2] # shape = (3,4), indices [2,5,8] x [1,...,4] x [2]
\end{CodeInput}
\end{CodeChunk}

In the \code{masspcf.random} submodule, we provide \code{noisy_cos} and \code{noisy_sin} for random function generation.\footnote{It is our intention to extend this over time to include user-defined functions, as well.} These can be used to generate random arrays of PCFs of arbitrary dimension.

For each PCF, we begin by generating a sequence of increasing $0 = t_0 < t_1 < \cdots < t_n$, for some chosen $n$, with each $t_i$ sampled uniformly in $[0,1]$. Then, for each $t_i$, the generated function takes the value
\begin{align*}
    f(t_i) = g(2 \pi t_i) + \varepsilon(t_i),
\end{align*}
where $g(t)$ is either $\cos(t)$ or $\sin(t)$, and $\varepsilon(t)$ is iid $\mathcal{N}(0, \sigma)$ noise, with standard deviation $\sigma=0.1$ as the default.

The mean of an array \code{A} across a dimension \code{dim} can be computed using \code{mpcf.mean(A, dim=dim)}.

In the following example, we generate a $2 \times 10$ matrix of random functions and take their means across the columns of the matrix. The result is displayed in the figure directly following the example code.
\begin{CodeChunk}
\begin{CodeInput}
from masspcf.random import noisy_sin, noisy_cos

M = 10 # Number of PCFs for each case
A = mpcf.zeros((2,M))

# Generate 'M' noisy sin/cos functions @ 100 resp. 15 time points each.
# Assign the sin(x) functions into the first row of 'A' and cos(x)
# into the second row.
A[0,:] = noisy_sin((M,), n_points=100)
A[1,:] = noisy_cos((M,), n_points=15)

fig, ax = plt.subplots(1, 1, figsize=(6,2))

# Plot individual noisy sin/cos functions
# masspcf can plot one-dimensional arrays (views) of PCFs in a single line
plotpcf(A[0,:], ax=ax, color='b', linewidth=0.5, alpha=0.4)
plotpcf(A[1,:], ax=ax, color='r', linewidth=0.5, alpha=0.4)

# Means across first axis of 'A'
Aavg = mpcf.mean(A, dim=1)

# Plot means
plotpcf(Aavg[0], ax=ax, color='b', linewidth=2, label='sin')
plotpcf(Aavg[1], ax=ax, color='r', linewidth=2, label='cos')

ax.set_xlabel('t [2 pi]')
ax.set_ylabel('f(t)')
ax.legend()
\end{CodeInput} 
\end{CodeChunk}
\begin{figure}[H]
    \includegraphics{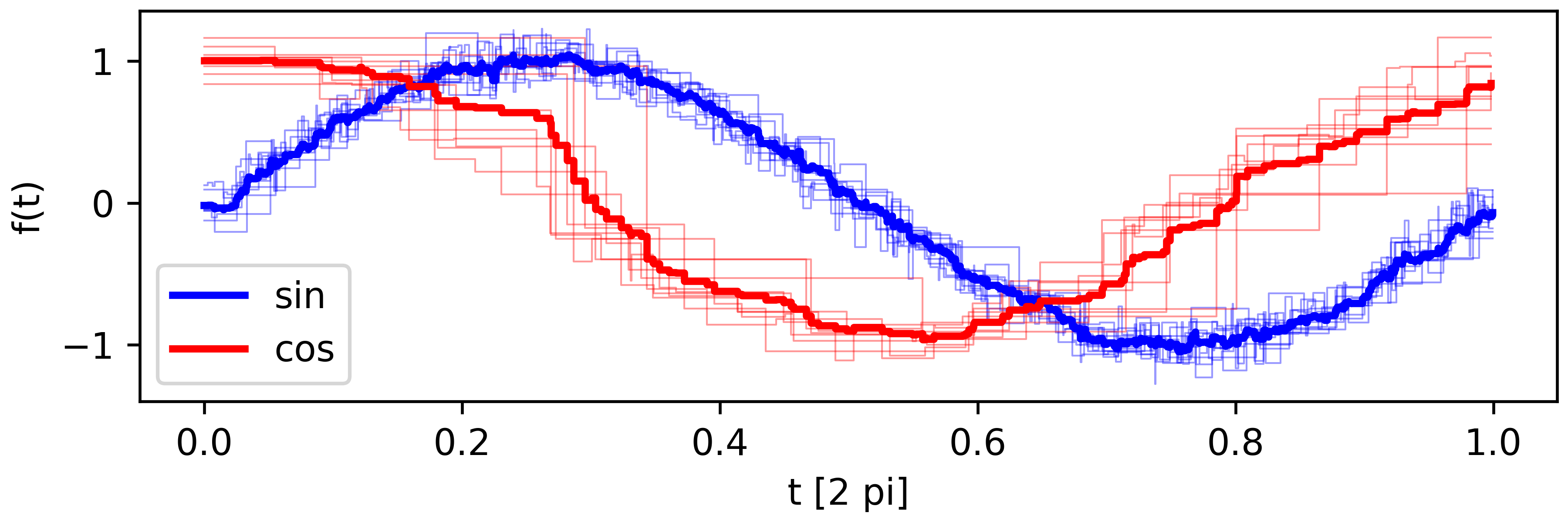}
\end{figure}

\subsection{Matrix computations}

We now turn our attention to computing distance and kernel matrices from a collection of PCFs. To this end, we begin by constructing a length 4 array of PCFs.
\begin{CodeChunk}
\begin{CodeInput}
f1 = mpcf.Pcf(np.array([[0., 5.], [2., 3.], [5., 0.]]))
f2 = mpcf.Pcf(np.array([[0., 2.], [4., 7.], [8., 1.], [9., 0.]]))
f3 = mpcf.Pcf(np.array([[0, 4], [2, 3], [3, 1], [5, 0]]))
f4 = mpcf.Pcf(np.array([[0, 2], [6, 1], [7, 0]]))

X = mpcf.Array([f1, f2, f3, f4])
\end{CodeInput}
\end{CodeChunk}

To obtain all pairwise distances between the functions in the array, we use the \code{pdist} command. By default, the $L_1$ distance is computed.
\begin{CodeChunk}
\begin{CodeInput}
mpcf.pdist(X)
\end{CodeInput}
\begin{CodeOutput}
[[ 0. 34.  6. 12.]
 [34.  0. 34. 24.]
 [ 6. 34.  0. 10.]
 [12. 24. 10.  0.]]
\end{CodeOutput}
\end{CodeChunk}
To use a different value of $p$ for the $L_p$ distance, we supply the optional argument \code{p}. As an example, we compute the $L_p$-distance with $p=3.5$.
\begin{CodeChunk}
\begin{CodeInput}
mpcf.pdist(X, p=3.5)
\end{CodeInput}
\begin{CodeOutput}
[[ 0.          9.80058139  2.49774585  3.81895602]
 [ 9.80058139  0.         10.10250875  8.76880217]
 [ 2.49774585 10.10250875  0.          2.82601424]
 [ 3.81895602  8.76880217  2.82601424  0.        ]]
\end{CodeOutput}
\end{CodeChunk}
Finally, to compute the $L_2$ Gram (inner product) matrix, we use the \code{l2_kernel} command.
\begin{CodeChunk}
\begin{CodeInput}
mpcf.l2_kernel(X)
\end{CodeInput}
\begin{CodeOutput}
[[ 77.  53.  55.  38.]
 [ 53. 213.  31.  51.]
 [ 55.  31.  43.  26.]
 [ 38.  51.  26.  25.]]    
\end{CodeOutput}
\end{CodeChunk}

\section{Implementation details and performance} \label{sec:implementation}

On the CPU side, we use \pkg{Taskflow} \citep{huang2021taskflow} for \proglang{C++} for parallelization. For the GPU implementation, we use a hand-written NVIDIA CUDA \citep{nickolls2008scalable, nvidiacuda} implementation.

For the distance computations, since the matrices we compute can be quite large, we have resorted to computing in block rows on the GPU. This has the added benefit that we can easily utilize several GPUs in a multi-GPU setting, by computing different blocks on different GPUs. These computations are scheduled using \pkg{Taskflow} and make use of its work-stealing  capabilities. 

Work stealing \citep{burton1981executing} is a scheduling paradigm in which a task that was originally scheduled to be executed on one compute unit (e.g., a GPU) can be repartitioned to---or \emph{stolen} by---another compute unit if it runs out of tasks while the compute unit on which the task originally was scheduled is still busy. We do not know \emph{a priori} how much work is required in each block row since this depends on the structure of each PCF pair in the computation, so having an automatic way of rescheduling work is crucial to achieving high performance.

The algorithms that return lists (Algorithms \ref{alg:iteratesegments} and \ref{alg:iteraterects}) have been implemented to instead invoke a callback for each rectangle/segment. This way, no extra allocations or memory transfers are necessary, and we can rely on the inlining capabilities of modern compilers to avoid extra function calls.

For long-running computations we use an asynchronous tasking model so that we can pass back control from \proglang{C++} to \proglang{Python} repeatedly during the execution. This has two benefits: 1) it makes the task interruptable by the user, and 2) we can display progress information and later provide progress hooks should the need arise.

The multidimensional arrays are implemented using \pkg{xtensor} \citep{xtensor} as its \proglang{C++} backend. We expose a limited set of the full functionality of \pkg{xtensor} as part of the \pkg{masspcf} package.\footnote{For example, we do not include matrix-matrix multiplication as this is not well-defined for PCFs.}

The \proglang{Python}/\proglang{C++} interface uses \pkg{pybind11} \citep{pybind11}.

\begin{figure}
     \centering
     \begin{subfigure}[T]{0.45\textwidth}
         \centering
         \includegraphics[width=\linewidth]{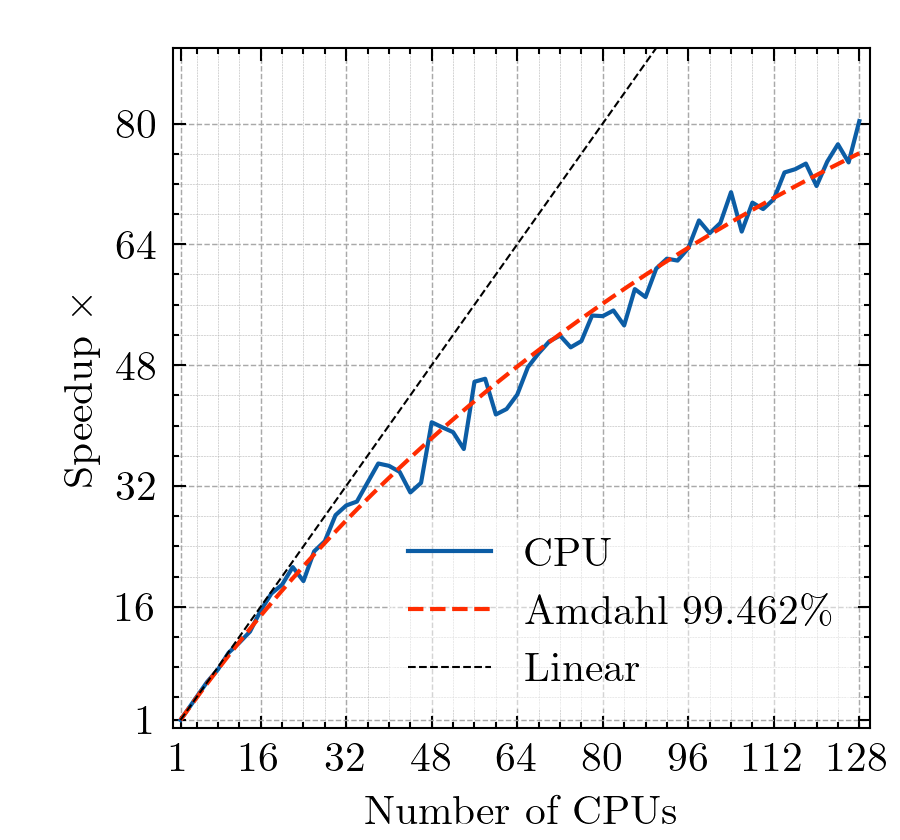}
         \caption{~}
         \label{fig:cpuscaling}
     \end{subfigure}
     \begin{subfigure}[T]{0.45\textwidth}
         \centering
         \includegraphics[width=\linewidth]{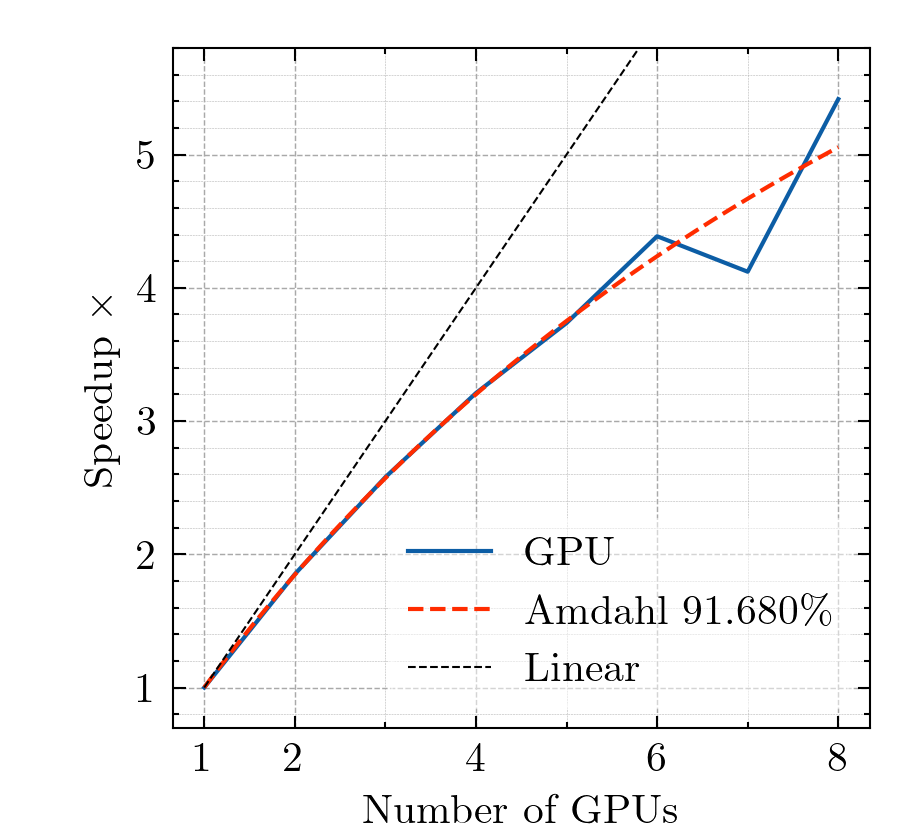}
         \caption{~}
         \label{fig:gpuscaling}
     \end{subfigure}%
        \caption{Speedup on synthetic dataset, running on (a) multiple CPUs, and (b) multiple GPUs. Times are averaged over 10 runs. The dataset is regenerated with a new random seed for each run. In addition, we plot fitted Amdahl curves and display the program parallel portion as a percentage. We used 2,500 PCFs for the CPU side and 50,000 PCFs when running on GPU.}
        \label{fig:scaling}
\end{figure}

Next, we discuss the performance and scalability of the software. For this, we first generated synthetic data following the procedure in Appendix \ref{app:synthetic}. The resulting dataset contains PCFs of varying length (10-1000 time points) and time scales. The operation tested is computing the $L_1$ distance matrix for the set of PCFs.

To test the scalability of our package to multiple CPUs/GPUs, we used a system with two AMD Epyc 7742 CPUs, each with 64 physical cores, and eight NVIDIA RTX A100 GPUs. We plot the speedups in Figure \ref{fig:scaling} together with Amdahl (see e.g., Ch 1 in \citet{hwu2022programming}) fits. We observe near-linear scaling up to 16 CPU cores and two GPUs, but there is substantial benefit in going all the way up to 128 CPU cores and eight GPUs. 

Given the Amdahl trajectories, we believe our software will continue to scale well for the foreseeable future. Indeed, we have computed the pairwise distances between 500,000 PCFs (125 billion integrals) using the 8-GPU configuration. This computation finished in around 423 seconds.

\begin{figure}
     \centering
     \begin{subfigure}[T]{0.45\textwidth}
         \centering
         \includegraphics[width=\linewidth]{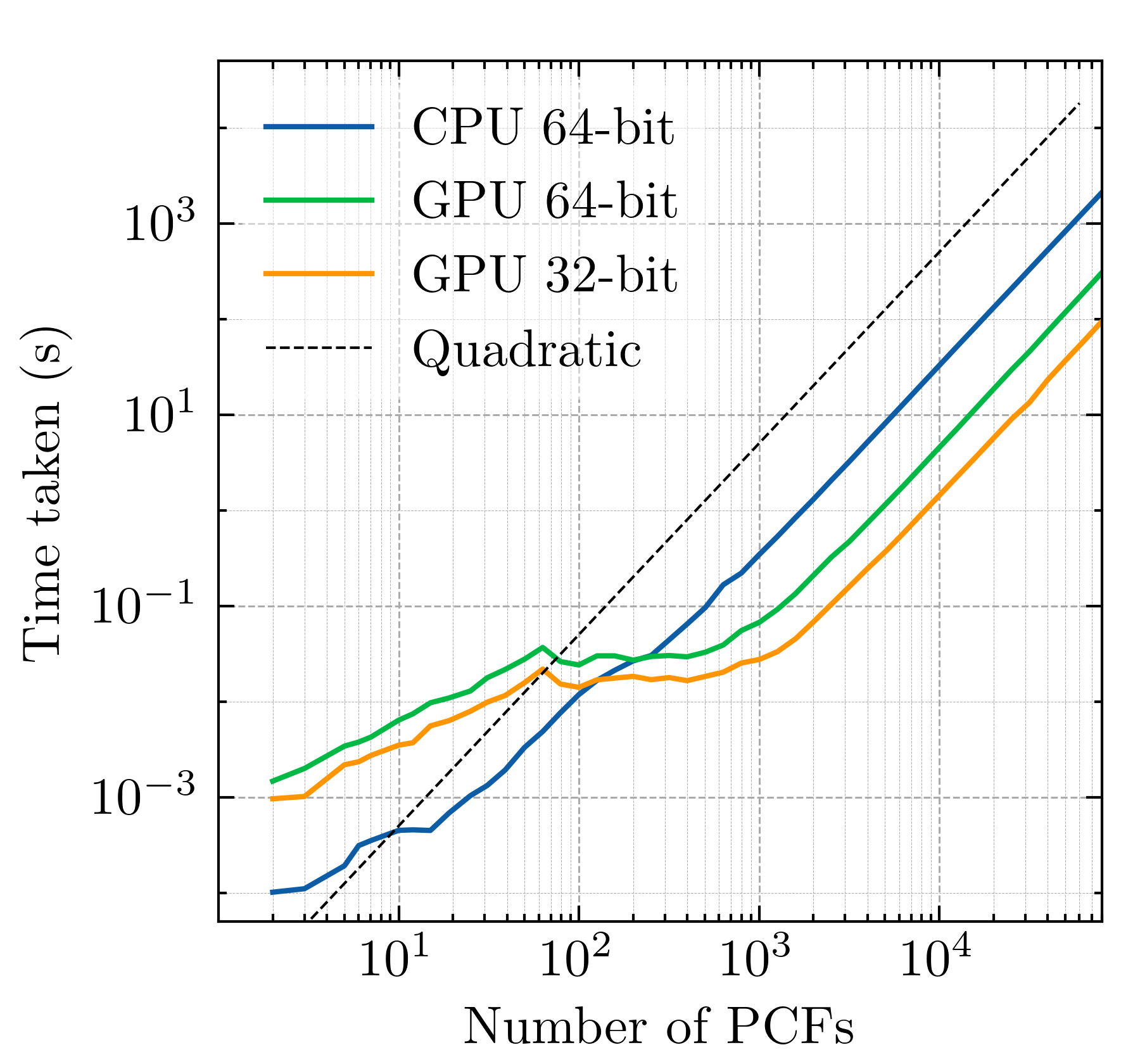}
         \caption{~}
         \label{fig:npts}
     \end{subfigure}
     \begin{subfigure}[T]{0.45\textwidth}
         \centering
         \includegraphics[width=\linewidth]{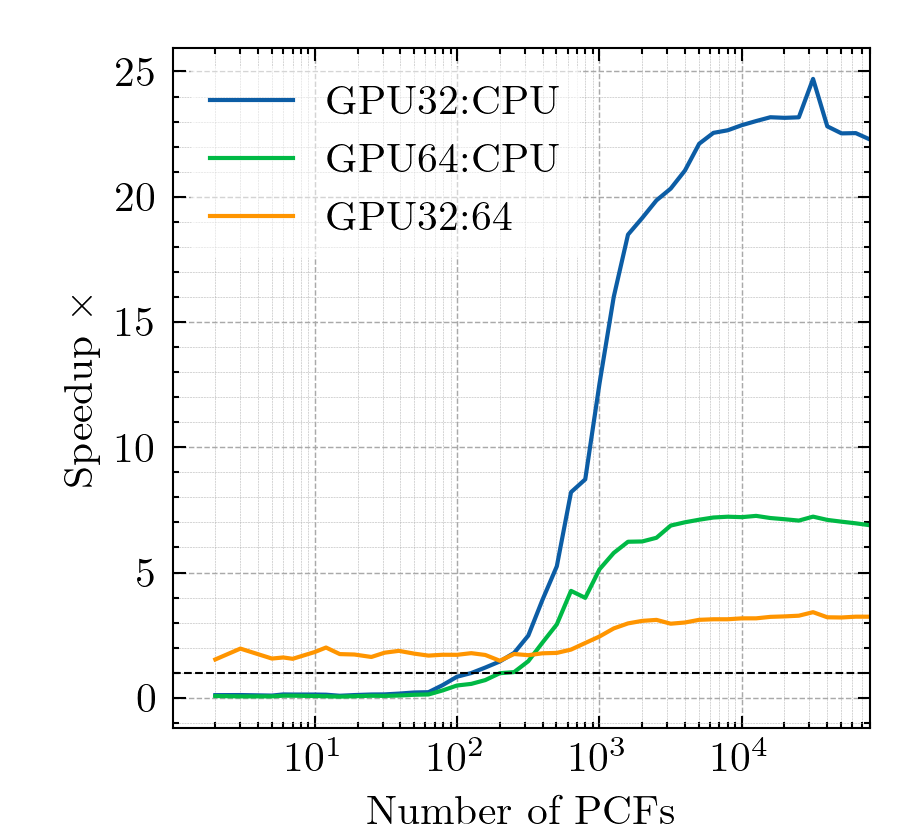}
         \caption{~}
         \label{fig:relative}
     \end{subfigure}%
        \caption{Benchmarks using different floating point precision (32 vs 64-bit) on CPU/GPU. We display wall running times (a) and speedups (b) for different numbers of PCFs. For $M<500$ PCFs, we display the mean of 10 runs for each $M$, and for $M \geq 500$, we instead use the mean over 3 runs. We see that for a small number of PCFs, the CPU implementation is faster, so in the implementation, we automatically switch to use CPU for small datasets. We also benchmarked using 32-bit floats on CPU but as expected the performance is nearly identical to the 64-bit case, so we do not include this benchmark in the figures. On GPU, using lower precision is significantly faster.}
        \label{fig:cpugpu}
\end{figure}

We stress, however, that it is \emph{not} necessary to use a supercomputer to run the software. To the contrary, we have used the package on a variety of consumer-grade desktop and laptop computers without issue.

To test how the software performs in a more typical workstation setup, we used eight cores of an Intel Xeon W-2295 CPU and one NVIDIA RTX A4000 GPU. We plot wall running times and speedups in Figure \ref{fig:cpugpu}. 

As expected, the CPU outperforms the GPU when the number of PCFs is relatively small. 
Additionally, we see the benefit of using 32-bit floating point numbers on the GPU side. We also tested using 32-bit floats when running on CPU, but as expected the difference from the 64-bit case is minuscule.

\section{Future work} \label{sec:future}

We believe that our methods could be extended relatively easily to work with piecewise \emph{linear} functions, such as persistence landscapes \citep{bubenik2015statistical} and clique-/facegrams \citep{dlotko2023combinatorial} (for the one-parameter case). We note that the derivative of a piecewise linear function, $f \colon [0,\infty) \to \mathbb{R}$, on each of its pieces is a PCF and that, together with the value of $f(0)$, this completely determines $f$.

Also, at the moment, the only compute accelerators we support are CUDA-capable GPUs. It may be interesting to develop the package further to support, e.g., Apple Metal and AMD ROCm.

\section*{Acknowledgments}

This work was partially supported by the Wallenberg AI, Autonomous Systems and Software Program (WASP) funded by the Knut and Alice Wallenberg Foundation.

We thank Wojciech Chach{\'o}lski and Gonzalo Uribarri for insightful discussions; Adam Breitholtz, Isaac Ren, Ricky Mol\'{e}n and Barbara Mahler for testing the \proglang{Python} module; Matt Timmermans for making us aware of the accumulator design for the parallel reductions; and Ryan Ramanujam for providing data that we used while developing the library.

The large-scale benchmarking was enabled by the Berzelius resource provided by the Knut and Alice Wallenberg Foundation at the National Supercomputer Centre.


\bibliography{refs}

\clearpage

\begin{appendix}

\section{Synthetic data generation}
\label{app:synthetic}

The synthetic data used in the experiments were generated in the following manner. Let $M$ be the number of PCFs to use. We repeat the following procedure for each of the PCFs.

\begin{enumerate}
    \item Draw a number $n$, from $\mathcal{U}([10,1000])$, a uniform integer distribution from 10 to 1000 with the endpoints included.
    \item Draw $\alpha$ from $\mathcal{N}(0,1)$, a standard normal distribution.
    \item Draw $n-1$ numbers $\tilde{t}_1,\ldots,\tilde{t}_{n-1}$ from $\mathcal{N}(0,1)$
    \item Draw $n-1$ numbers $v_0,\ldots,v_{n-2}$ from $\mathcal{N}(0,1)$.
    \item Sort $\tilde{t}_1,\ldots,\tilde{t}_{n-1}$ in increasing order by magnitude to form a new list $t_1,\ldots,t_{n-1}$.
    \item Construct the PCF from the following matrix:
    \begin{align*}
        \begin{bmatrix}
            0 & \alpha t_1 & \cdots & \alpha t_{n-2} & \alpha t_{n-1} \\
            v_0 & v_1 & \cdots & v_{n-2} & 0
        \end{bmatrix}^T.
    \end{align*}
\end{enumerate}

This generates an arbitrary number of PCFs that have
\begin{itemize}
    \item different number of points,
    \item different time scales (due to $\alpha$), and
    \item final value equal to 0 (so that integrals over $[0,\infty)$ are finite).
\end{itemize}

\begin{figure}
    \centering
    \includegraphics[width=0.8\linewidth]{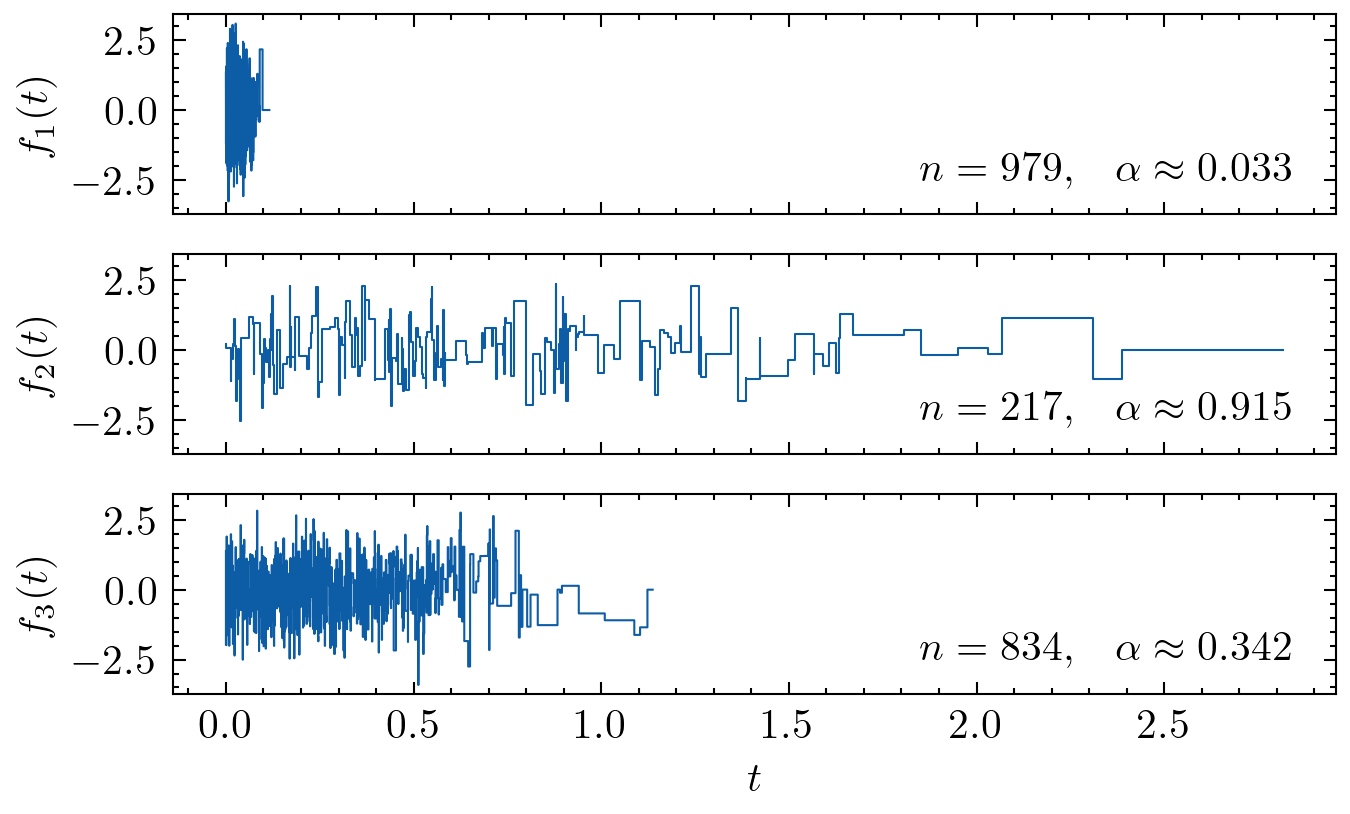}
    \caption{A representative sample of three PCFs generated from the synthetic data generation procedure described in the appendix. Here, $f_1$, $f_2$ and $f_3$ have 979, 217 and 834 time points, respectively.}
    \label{fig:panel3xgen}
\end{figure}

We plot a representative sample of PCFs generated from this procedure in Figure \ref{fig:panel3xgen}.

\end{appendix}

\end{document}